\documentclass[useAMS,usenatbib]{mn2e}

\usepackage{lscape}
\usepackage{longtable}

\usepackage{graphicx}

\title[II. Apples to apples $A^2$: selection functions for next-generation surveys]{II. Apples to apples $A^2$: cluster selection functions for next-generation surveys}
\author[Ascaso, Mei, Bartlett \& Ben\'itez]{B. Ascaso$^{1,2}$ \thanks{E-mail: ascaso@apc.univ-paris7.fr};  S. Mei$^{2,3,4}$, J. G. Bartlett$^{1}$, N. Ben\'itez$^{5}$\\
$^{1}$
APC, AstroParticule et Cosmologie, Universit\'e Paris Diderot, CNRS/IN2P3, CEA/lrfu, Observatoire de Paris, Sorbonne Paris Cit\'e, \\
10, rue Alice Domon et L\'eonie Duquet, 75205 Paris Cedex 13, France\\
$^{2}$GEPI, Observatoire de Paris, PSL Research University,  CNRS, University of Paris    Diderot,\\
 61, Avenue de l'Observatoire 75014, Paris France \\
$^{3}$University of Paris Denis Diderot, University of Paris Sorbonne Cit\'e (PSC), 75205 Paris Cedex
  13, France\\
$^{4}$California Institute of Technology, Pasadena, CA 91125, USA\\
$^{5}$Instituto de Astrof\'isica de Andaluc\'ia (IAA-CSIC), Glorieta de la Astronom\'ia s/n, 18008, Granada, Spain\\}

\begin{document}

\date{Accepted . Received }


\maketitle

\label{firstpage}

\begin{abstract}

We present the cluster selection function for three of the largest next-generation stage-IV surveys in the optical and infrared: Euclid-Optimistic, Euclid-Pessimistic and the Large Synoptic Survey Telescope (LSST). To simulate these surveys, we use the realistic mock catalogues introduced in the first paper of this series.

We detected galaxy clusters using the Bayesian Cluster Finder (BCF) in the mock catalogues. We then modeled and calibrated the total cluster stellar mass observable-theoretical mass ($M^*_{\rm CL}-M_{\rm h}$) relation using a power law model, including a possible redshift evolution term. We find a moderate scatter of $\sigma_{M^*_{\rm CL} | M_{\rm h}}$ of 0.124, 0.135 and 0.136 $\rm dex$ for Euclid-Optimistic, Euclid-Pessimistic and LSST, respectively, comparable to other work over more limited ranges of redshift.  Moreover, the three datasets are consistent with negligible evolution with redshift, in agreement with observational and simulation results in the literature.

We find that Euclid-Optimistic will be able to detect clusters with $>80\%$ completeness and purity down to $8\times10^{13} h^{-1} M_{\odot}$ up to $z<1$. At higher redshifts, the same completeness and purity are obtained with the larger mass threshold of $2\times10^{14} h^{-1} M_{\odot}$ up to $z=2$. The Euclid-Pessimistic selection function has a similar shape with $\sim10\%$ higher mass limit. LSST shows  $\sim 5\%$ higher  mass limit than Euclid-Optimistic up to $z<0.7$ and increases afterwards, reaching values of $2\times10^{14} h^{-1} M_{\odot}$ at $z=1.4$.  Similar selection functions with only $80\%$ completeness threshold have been also computed. The complementarity of these results with selection functions for surveys in other bands is discussed.

\end{abstract}


\section{Introduction}

At present, a large part of the extragalactic community is devoted to predicting the performance and limitations of the next-generation surveys by  analyzing a variety of simulations. The main goal of many of these surveys, for instance J-PAS \citep{benitez14}, Euclid \citep{laureijs11} and Large Synoptic Survey Telescope \citep[LSST, ][]{ivezic08,lsst09} among others, is the determination of the nature of dark energy. These surveys, in addition to unprecedented cosmological results, will bring enormous quantities of data to exploit and analyze. 

Galaxy clusters, the largest structures gravitationally bound in the Universe, are useful objects for the determination of the cosmological parameters  (e.g. \citealt{allen11}) as well as for the analysis of their galaxy population across time, (e.g. \citealt{ascaso08,ascaso09}). At present, several hundreds of thousands of structures up to moderate redshift ($z\sim 0.6$) have been censed in wide optical surveys including the Sloan Digital Sky Survey \citep[SDSS, ][]{koester07,hao10,szabo11,wen12,rykoff14}, the Canada France Hawaii Telescope Legacy Survey  \citep[CFHTLS, ][]{thanjavur09,adami10,milkeraitis10,durret11,ascaso12,licitra16}, the DLS  \citep{ascaso14a}, the Advanced Large, Homogeneous Area Medium Band Redshift Astronomical survey \citep[ALHAMBRA, ][]{ascaso15a} and a few hundred up to higher redshift ($z\sim 1.6$) in infrared surveys, such as the {\em Spitzer} IRAC Shallow Survey \citep{eisenhardt08} and the  {\em Spitzer} Wide-Area Infrared Extragalactic survey \citep[SWIRE, ][]{wen11},  the  {\em Spitzer} SPT Deep Field \citep[SSDF, ][]{rettura14}, the  {\em Spitzer} Adaptation of the Red-sequence Cluster Survey  \citep[SpARCS, ][]{muzzin08}, the Clusters Around Radio-Loud AGN program  \citep[CARLA, ][]{galametz12}.  In a few years, these numbers are expected to increase by a factor of  at least 10 (see \citealt{weinberg13}), with the advent of the next-generation surveys. Before handling such large quantities of data, we need to predict the kind of structures that each survey will be able to detect or, in the other words, their selection function.

Modeling and understanding accurately a selection function is not only important for providing a census of the properties of clusters and groups. It will also play a crucial role in constraining cosmological parameters using galaxy clusters as probes (e.g. \citealt{lima05,mantz08,cunha09,rozo10,mantz10,sartoris16}). In this sense, we need not only to know which clusters we will be able to detect with high completeness (ratio of the number of simulated clusters matched to a detected cluster to the total number of simulated clusters) and purity (ratio of the number of detected clusters matched to a simulated cluster to the total number of detected clusters) rates but, in addition, we will need to identify a mass proxy that allows us to determine halo mass with the best possible accuracy given the quality of the data. The introduction of a realistic scaling relation and its uncertainties are key to providing reliable cosmological constraints with galaxy cluster counts.

Up to date, the expected cluster selection function of many of the next-generation surveys in the optical is unknown. \cite{sartoris16} provided a cluster selection function for Euclid based on a pure analytical approach, whereas \cite{ascaso16} delivered a selection function of clusters and groups in the J-PAS survey by performing an empirical detection of clusters and groups in cosmological simulations. In this paper, we provided, for the first time, selection functions for the Euclid and LSST surveys using the latter empirical approach. While the cited surveys have not started yet, we are forced to work with mock catalogues that are known to be a fairly, though not perfect, representation of the reality (e.g. \citealt{ascaso15b}).

Different attempts have been made to define observables  and accurately calibrate  scaling relations at different wavelengths: the average X-ray temperature, $T_{\rm X}$, and luminosity, $L_{\rm X}$, for X-ray measurements, the total integrated Sunyav-Zel'dovich (SZ) signal over the cluster, $Y_{\rm SZ}$, for the SZ effect and shear and magnification for the weak lensing (WL) effect \citep{reiprich02,rozo11,rozo12,giodini13,rozo14a,rozo14b,anderson15}. However, little work has calibrated this relation with optical data, restricted to low-z regimes  \citep{yee03,rozo09,andreon10,vanuitert16,ascaso16} or to few clusters up to higher redshift \citep{andreon12,saro15,licitra16}. This  paper work calibrates, for the first time, scaling relations using very large samples of optical clusters both in mass and redshift and demonstrates the ability of optical data to obtain robust constraints for this relation that can be competitive with other techniques, such as X-rays, SZ or WL.

This paper is the second in a series entitled `Apples to Apples'  (\emph{$A^2$})  that aims to compare galaxy cluster features (photometric properties, selection functions, observable-theoretical mass relation, cosmological constraints) for different next-generations surveys, using the same mock catalogues and methodology. The first paper of the series (\citealt{ascaso15b}, hereafter  \emph{$A^2$}I) introduced the mock catalogues and characterized the photometry and photometric redshift performance, checking that the properties of the galaxies resembled those of real data for two next generation surveys: LSST and Euclid.

In this paper, we first detect galaxy clusters using the Bayesian Cluster Finder (BCF, \citealt{ascaso12,ascaso14a,ascaso15a}), a code to detect galaxy clusters and groups in the optical even with the absence of a red sequence by using a Bayesian variation of the matched filter technique, in the different mock catalogues already tested in \emph{$A^2$}I. Then, we model the cluster selection function, finding the minimum mass limit for which we can reliably detect galaxy clusters with completeness and purity rates higher than a given percentage, and we provide the observable-theoretical mass relation for the same surveys, using a consistent methodology. In future work of the  \emph{$A^2$} series (Ascaso et al. in prep), we will use the selection function and observable-mass relation determined in this work to obtain reliable cosmological constraints from cluster counts for the two next-generations surveys considered in the series. This way, we will complete a consistent comparison of the performance in handling galaxy clusters and groups for LSST and two different Euclid surveys.

The structure of the paper is as follows. In section 2, we provide a short description of the different next-generation surveys considered in this series. Section 3  describes the procedure performed to obtained the mock catalogues. Section 4 summarizes the main features of the Bayesian Cluster Finder (BCF) applied to the mocks. In Section 5 we first show the completeness and purity rates obtained for the different surveys, then fit a model for the observable-theoretical mass relation and compare the results, and finally obtain the selection function and compare it with other surveys. Section 6 summarizes our conclusion.

The cosmology used throughout this paper is of $\Omega_{\rm M}=0.25$, $\Omega_{\Lambda}=0.75$, $\Omega_{\rm b}=0.045$, $\sigma_{\rm 8}=0.9$, and $n_{\rm s}=1$ and $h=0.73$ in order to be consistent with that of the mock catalogues. All the magnitudes in the paper are given in the $AB$ system and all the halo masses are expressed in units of $h^{-1} M_{\odot}$. Throughout this paper, we will refer to $\log$ as the decimal logarithm.

\section{Considered next generation surveys}

In this series, we have focused on two next-generation surveys:  LSST and Euclid. A detailed description of the mocks that mimic these surveys is given in \emph{$A^2$}I and in the canonical papers of each of the surveys (see below). In this section, we give a brief description of the two surveys considered. 

\subsection{LSST}

The Large Synoptic Survey Telescope (LSST,  \citealt{ivezic08,lsst09}) will begin taking data in 2020. The main objective of this survey is determination of the nature of dark energy by using different probes. Other topics, such as investigating the dark matter in the universe, studying galaxy evolution, exploiting the transients or imaging deeply the Milky Way, are also among the priorities of the LSST.

LSST will collect data from an 8.4m telescope placed on Cerro Pach\'on (Chile). The survey will cover 18000 deg$^2$ in six broad optical bands, $ugrizY$, to a depth of $r=27.5$ mag at the end of the survey. 

In \emph{$A^2$}I, we used \texttt{PhotReal} (\citealt{ascaso15b}, Ben\'itez et al. in prep) to simulate mock catalogues down to the depth prescribed in Table 1 of \cite{ivezic08}. The photometric errors were estimated following the prescription given by the LSST Survey Science Group. We verified that the photometry and photometric errors resemble those as the real clusters.

\subsection{Euclid}

The Euclid survey \citep{laureijs11} is a European space mission, with a starting date planned for 2020. While the main goal of the Euclid survey is  understanding of dark energy using different probes, a host of ancillary legacy science goals have also been defined, ranging from galaxy evolution to stellar physics.

Euclid plans to complete a `Wide' and a `Deep' survey, the latter being two magnitudes deeper than the former. The size of the `Deep' survey will only be 40 square degrees, whereas the `Wide' survey will cover 15,000 square degrees. In this work, since we are mostly interested in galaxy clusters, we have focused on the ÃWideÃ survey to have enough area to detect a statistically significant number of these rare structures.

The entire area will be imaged in three infrared bands, $YJH$, down to $H\sim24$ mag. Moreover, near-infrared spectroscopy will be obtained for the brightest $H\alpha$ objects.  In \emph{$A^2$}I, we generated photometric errors for the Euclid bands by modeling the photometric errors in existing data from the Cosmic Assembly Near-infrared Deep Extragalactic Legacy Survey (CANDELS\footnote{http://candels.ucolick.org/},  \citealt{guo13b}), re-normalizing to the Euclid depth.

Since Euclid is an IR survey, it is planned to be combined with different sets of ground-based observations. In this series, we have considered two scenarios that, following the nomenclature introduced in \emph{$A^2$}I, are:

\begin{itemize}
\item Euclid pessimistic case:  the optical complement will consist of broad-band photometry in five bands ($grizY$) from the DES survey down to the depth stated in Table 1 of \cite{mohr12}.  Photometric errors for the DES bands have been estimated from the mock catalogues by \cite{chang14}.
\item Euclid optimistic case:  the optical data will come from the same five broad-bands ($grizY$) of the DES survey as in the pessimistic case, plus the six broad optical bands ($ugrizY$) from the LSST survey as described in section 2.1.
\end{itemize}

\section{Mock catalogues}

The main mock catalogues used in this series of work are fully explained in \emph{$A^2$}I and publicly available to the community\footnote{http://photmocks.obspm.fr/}. We briefly summarize here the procedure to build them and refer the reader to \emph{$A^2$}I for further details.

These mock catalogues are meant to be as realistic as possible since our aim is to obtain accurate (and therefore, realistic) selection function estimates. For this reason, we first considered the  500 deg$^2$ Euclid mock catalogues by \cite{merson13}\footnote{http://community.dur.ac.uk/a.i.merson/lightcones.html}. These mocks are built from the dark matter halo merger trees extracted from a $2160^3$-particle cube N-body Millennium Simulation \citep{springel05}. Then, the haloes were populated with galaxies created with the semi-analytical model GALFORM  \citep{cole00,bower06}. 

After close inspection of the colors of the galaxies, we realized that some cluster properties, such as the color-magnitude relation, photometric redshifts or photometric errors, did not match those of real data  (see \emph{$A^2$}I for a battery of tests). We applied  \texttt{PhotReal} (\citealt{ascaso15b}, Ben\'itez et al. 2016, in prep) to these mocks in order to create more realistic galaxy photometric sets by computing a new photometry set from a well-calibrated library of spectra.

The final mock includes the realistic \texttt{PhotReal} photometry and photometric errors, their associated photometric redshifts and the dark matter halo information originally from the N-body Millennium simulation. As shown in \emph{$A^2$}I, the photometric properties of these galaxies resemble those of real data. For instance, the color-magnitude relation reproduced that observed in real clusters and other global properties of the galaxies, such as the stellar mass function, the luminosity function and the angular function both for quiescent and star-forming galaxies, also agreed with observations.

Only haloes with masses higher than $3\times10^{13} h^{-1} M_{\odot}$ are treated as clusters and groups in this work. The number of these haloes in the mock catalogue amount to 72329, up to $z\le 2$. For reference, the number of haloes more massive than $10^{14} h^{-1} M_{\odot}$ are 7270, up to the same redshift. This number is large enough to obtain reasonable statistics at the high-mass end.

\section{The Bayesian Cluster Finder}

We used the Bayesian Cluster Finder, (BCF, \citealt{ascaso12,ascaso14a,ascaso15a}) to detect galaxy clusters and groups in the Euclid and LSST mock catalogues described in  section 3.  We have chosen this cluster detector for two main reasons. First, it does not depend on the presence or absence of the red sequence, which makes it particularly suitable for detecting galaxy clusters at high redshift. Second, the BCF has already been used in a variety of surveys with reliable results. In particular, a minimum 70\% agreement has been found when comparing the detections with other optical, X-ray and SZ sets in different studies: the CFHTLS-Archive Research Survey \citep{ascaso12}; the Deep Lens Survey \citep{ascaso14a} and the ALHAMBRA survey \citep{ascaso15a}. Furthermore, it has also been applied to a mock catalogue mimicking the J-PAS survey \citep{ascaso16}  to obtain the survey selection function (see section 5.3 for a detailed comparison between the different selection functions obtained for different next-generation surveys). While the BCF has been described in the original publications, we give here a brief summary of the main performance of the algorithm and refer the reader to the original publications for more information.

The BCF is the first Bayesian algorithm built to detect galaxy clusters and groups with optical and IR data. The algorithm first calculates, for each galaxy in the survey, the probability that there is a cluster centered on it at a given redshift slice. This probability is calculated following a Bayesian prescription. For the likelihood, we model and convolve the luminosity function, density profile and photometric redshift distribution of the cluster  to obtain the likelihood. In addition, we assume different priors that model properties of the clusters that are not necessarily always present. Their function is hence to enhance the posterior probability without penalizing the detection for not showing a particular property. In our case, we model the presence of a red sequence of galaxies at a particular redshift of the clusters and the relation of the magnitude of the Brightest Cluster Galaxy (BCG) to the redshift of the cluster. The red sequence is modeled by computing the expected colors at different redshifts for an early-type passive elliptical template from \cite{coleman80} and assuming a fixed slope obtained from a sample of well-characterized galaxy clusters. The main colors used, chosen to sample the 4000\AA\, break efficiently, are $(g-i)$ for $z<0.9$, $(i-z)$ for $0.9<z<1.4$. For the two Euclid surveys, we also used the color $(z-H)$ for redshift slices at $z>1.4$. The masks are taken into account, as well as the photometric redshift resolution of the survey. Indeed, the redshift slices considered for this work are separated by a bin width of 0.1, which is 2-3 times the photometric dispersion of the surveys considered.

Once we have computed the probabilities, we search for peaks in the probability density maps, and we select as clusters those peaks above $3\sigma$, where $\sigma$ is the scatter of the background galaxies. Spatially contiguous galaxies are associated to the same peak. We start from the peak with the highest signal and, following an iterative process, we end when no galaxies over the threshold are left. Finally, those detections separated by less than 0.5 Mpc in projected space and separated up to two bins in redshift space, are merged into one.

The final output of the algorithm provides a list of clusters with the coordinates of their central galaxy where the probability reaches its maximum, an estimation of their redshift obtained from fitting a Gaussian to the photometric redshifts of the galaxies statistically belonging to the cluster, and a measurement of their richness or mass observable. In this work, we consider the  total stellar mass, $M^*_{\rm CL}$, as in previous studies \citep{ascaso15a,ascaso16}. This quantity computes the sum of the stellar mass of all the galaxies statistically belonging to the clusters brighter than a magnitude limit that depends on the depth of the survey within a certain radius. Since we wanted to compare the different measurements, we have considered a limit of absolute $i$-band magnitude, $M_{\rm i}$, of -21 mag. This magnitude limit ensures that we do not introduce any bias in the mass measurement since the luminosity function that we are sampling for all the different surveys is complete at least until redshift $z<1.5$. The radius has been optimized following the same approach as in \cite{ascaso16}. We  performed a test where we compute the $M^*_{\rm CL}$ for different radii ranging from 0.5 to 1.5 Mpc in steps of 0.25 Mpc, and we compute the scatter in the observable-theoretical mass relation (see section 5.1). Then, we have chosen the aperture to compute $M^*_{\rm CL}$ that minimizes the dispersion, which in this case turns out to be 1 Mpc.

The $M^*_{\rm CL}$ has proven  to be a good cluster mass proxy in surveys with very high photometric redshift accuracy (e.g. the ALHAMBRA survey, \citealt{ascaso15a}, or the J-PAS survey, \citealt{ascaso16}) due to the low percentage of outliers derived from this multi-band photometric data. In surveys with lower photometric redshift resolution, such as the LSST or Euclid \citep{ascaso15b}, the computation of $M^*_{\rm CL}$ might be biased at higher redshifts. We will take this issue into account in a forecoming paper (Ascaso et al. in prep), where we will perform a detailed study on comparing different observables or cluster mass proxies.

We have run the BCF on the three mock catalogues introduced in section section 2, and we have matched the original mock sample to the recovered sample following the same Friends-of-Friends (FoF, \citealt{huchra82}) algorithm described in \cite{ascaso12,ascaso14a,ascaso15a,ascaso16}. Basically, we consider a detection to be a `friend' to an original detection if their centres are separated less than 3 Mpc in angular comoving distance, including errors. We start by building the list of friends of friends for each candidate until no more friends can be added. Then, the `friend' with the closest photometric redshift to the original detection is selected. If this detection is found to be less than 1 Mpc, we keep this detection as a match. Otherwise, we discard it.

\section{Clusters in next generation surveys}

In this section, we aim to characterize the selection function of the clusters in the three considered surveys. We first fit the observable-mass relation to the matched clusters and compute an estimation of the total masses of the clusters (section 5.1). Second, we compute the completeness and purity rates using this calibration  to express the purity as a function of halo mass and obtain the minimum mass threshold to reach both very conservative ($>80\%$) completeness and purity rates  (section 5.2). We proceed to fit again the  observable-mass relation down to these halo mass limits and iterate this procedure until it converges.  Then we compute the final selection function in section 5.3 and compare it to those obtained for  other surveys. Finally, we explore the redshift accuracy obtained for the clusters detected (section 5.4).

\subsection{Observable-Mass Relation}

Understanding and controlling uncertainties in the translation of an observable into a theoretical quantity is of great importance, in order not to misestimate the final results that are derived from these quantities. In our case, we are interested in knowing the cluster total mass, $M_{\rm h}$, which is not a direct observable. Instead, we have a measurement of the total stellar mass, $M^*_{\rm CL}$. Therefore, accurately determining the relation of this observable to the mass, $M^*_{\rm CL}-M_{\rm h}$, and its scatter, is key to obtaining cosmological predictions (e.g. \citealt{rozo10} and references therein).

As mentioned before, $M^*_{\rm CL}$ has proven to be robust in other work dealing with very high resolution photometric redshift data (e.g. ALHAMBRA and J-PAS survey). The fact that the rate of photometric redshift outliers is very low ($<1-2 \%$) for these surveys allows us to include in the computation of   $M^*_{\rm CL}$  galaxies belonging to the clusters with a very little contamination from outliers. For the surveys considered here, the expected photometric redshift scatter is significantly higher  (see \citealt{ascaso15b} for a consistent comparison of the photometric redshift properties of the different surveys). Therefore, some biases could be introduced in the computation of  $M^*_{\rm CL}$, particularly at high redshift.  A future paper will be devoted to designing and exploring different observables to select those that minimize the scatter.

\begin{figure*}
\centering
\includegraphics[clip,angle=90,width=1.0\hsize]{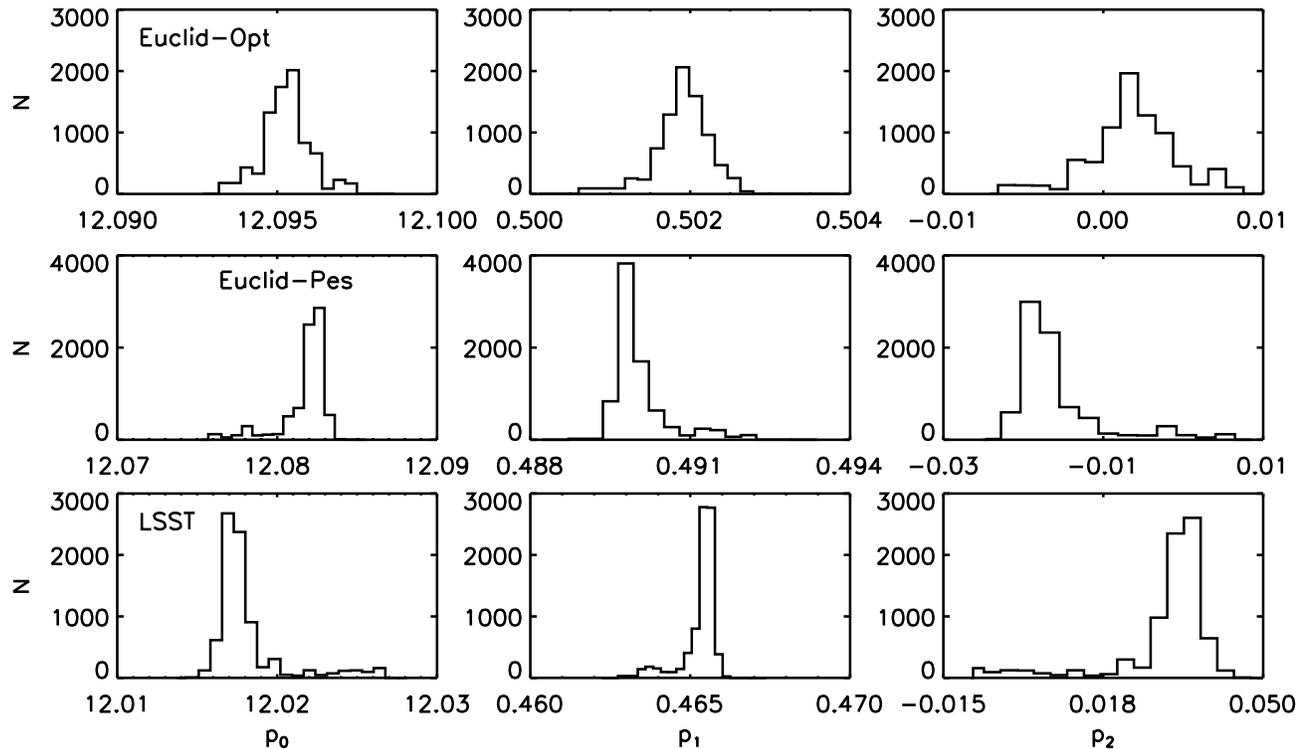} 
\caption{Distribution of the parameter fits for the observable-mass relation (equation \ref{eq:fitMM}) for the Euclid-Optimistic (upper row), Euclid-Pessimistic (middle row) and the LSST  (bottom row) surveys. These distributions were obtained from 8000 Monte Carlo realizations of the fit to the model described in equation  (\ref{eq:fitMM}). The results of the fit are collected in Table \ref{tab:fitMM}.}
\label{fig:fitParams}
\end{figure*}

The relation between the chosen observable and the theoretical mass is unknown. A number of studies (e.g. \citealt{lima05,lima07,rozo07,cunha09}) have proposed different models to fit this relation, mostly based on a linear relation in log-space between both variables. Some have also introduced a linear dependence in log-space with the redshift. Here, we test a model similar to that proposed by \cite{lima05,lima07,ascaso16}. Specifically, we choose to model the $M^*_{\rm CL}-M_{\rm h}$ relation as

\begin{equation}
\langle \log M^*_{\rm CL} | M_{\rm h},z \rangle=p_0+p_1\log \big( \frac{M_{\rm h}}{M_{\rm pivot}}\big)+ p_2 \log (1+z),
\label{eq:fitMM}
\end{equation}
where we have considered a power law relation between the two variables with a log dependence also on the redshift of the cluster. The $M_{\rm pivot}$ has been chosen to be $1\times 10^{14}M_{\odot}$ as an average value of the cluster sample. 

In this case, the $M_{\rm h}$ values are provided by the mock catalogue as the masses of the Dhaloes; these are regroupings of FoF sub-haloes found by the sub-routine \texttt{SUBFIND} (\cite{springel01}, see \cite{merson13,jiang14} for a more detailed explanation). \cite{jiang14} demonstrated that these detections were indistinguishable from the FoF detections defined by the linking length parameter $b=0.2$ for haloes more massive than $10^{12}M_{\odot}$. Moreover, these authors also showed that the mass of these haloes correlates with the virial mass $M_{200}$ better than the FoF masses. We fit this model for the cluster candidates detected with BCF restricted to $M_{\rm h} \ge 7\times 10^{13} h^{-1} M_{\odot}$ with both completeness and purity rates $>80\%$. This mass limit was determined as the minimum mass threshold obtained from computing both completeness and purity rates in section 5.2. The fit was performed using the well-known iterative non-linear  least-squares minimization method of Levenberg-Marquardt \citep{press92} as in \cite{ascaso16}. We performed 8000 Monte Carlo realizations of the fit, sampling a large range of possible initial conditions. The best fitting parameters for the model, together with their 68\% confidence level, are listed in Table \ref{tab:fitMM}.  Fig. \ref{fig:fitParams} shows the posterior probability of the fit parameters for all three surveys.

\begin{table}
      \caption{Best fitting parameters of the function (\ref{eq:fitMM}) together with their 68\% confidence level for the three different surveys considered in this work.}
      \[
         \begin{array}{cccc}
		\hline
		{\rm Parameter} &  {\rm Euclid-Opt}   &  {\rm Euclid-Pes} &  {\rm LSST}   \\\hline
p_0 &  0.08 \pm 0.002 & 0.08 \pm 0.004  & 0.08 \pm 0.005  \\
p_1 & 0.502 \pm 0.006 & 0.490 \pm 0.006   & 0.466 \pm 0.007 \\
p_2 & 0.002 \pm 0.001 & -0.018 \pm 0.011   & 0.034 \pm 0.021 \\
\hline
\sigma_{M^*_{\rm CL} | M_{\rm h},z} & 0.124 \, \rm dex & 0.135\, \rm dex & 0.136\, \rm dex\\ \hline 
	\end{array}
      \]
\label{tab:fitMM}
   \end{table}

\begin{figure}
\centering
\includegraphics[clip,angle=0,width=1.0\hsize]{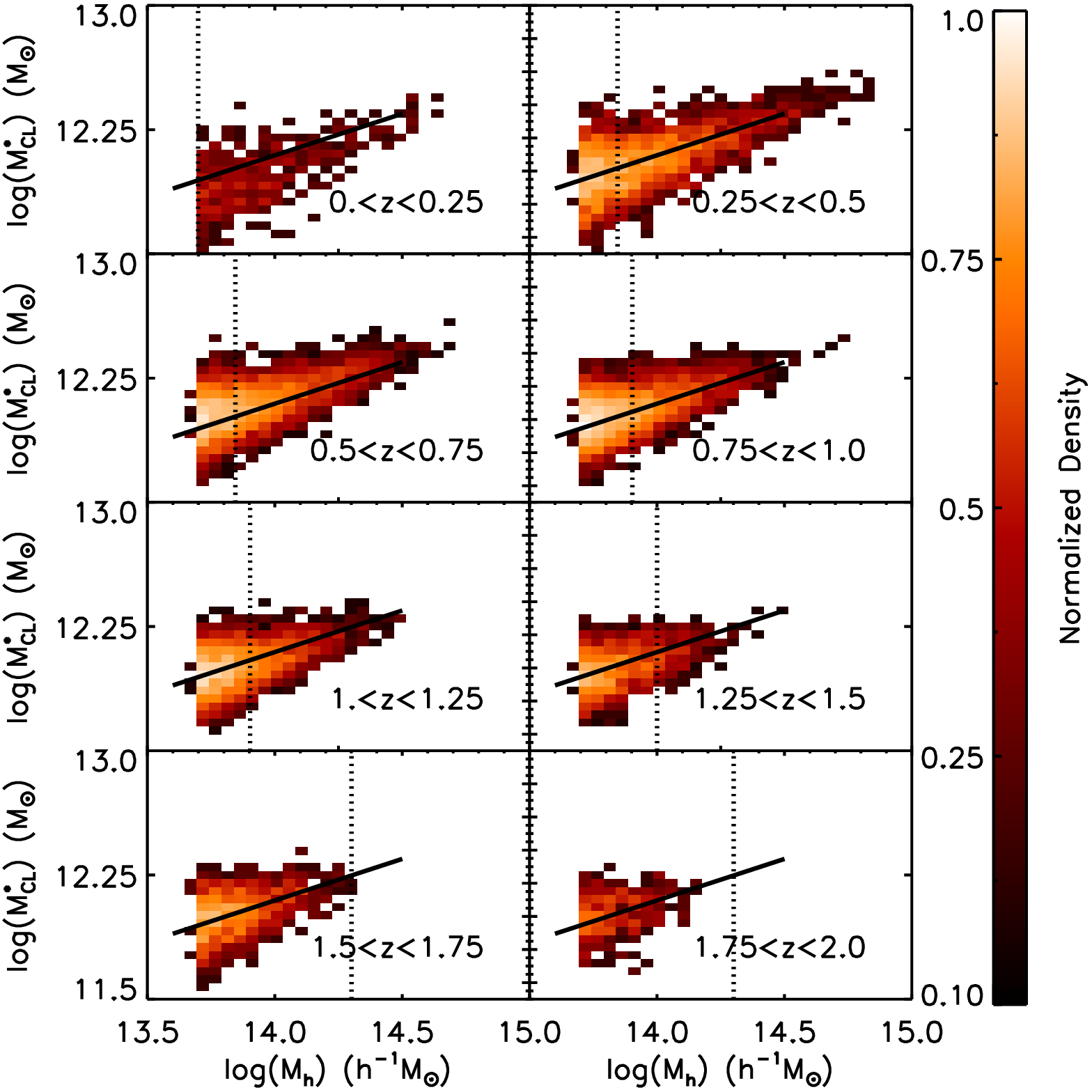} 
\caption{Density plots of the logarithm of the total stellar mass in the cluster as a function of the logarithm of the dark matter halo mass for the matched clusters in the Euclid-Optimistic mock catalogue for different redshift bins. The solid line indicates the linear fit obtained down to $M_{\rm h}=7\times10^{13}  h^{-1} M_{\odot}$ for the entire redshift range. The vertical dotted line refers to the mass limit for which we can reliable detect galaxy clusters based on the analysis performed in section 5.2.}
\label{fig:smassmhaloEO}
\end{figure}

\begin{figure}
\centering
\includegraphics[clip,angle=0,width=1.0\hsize]{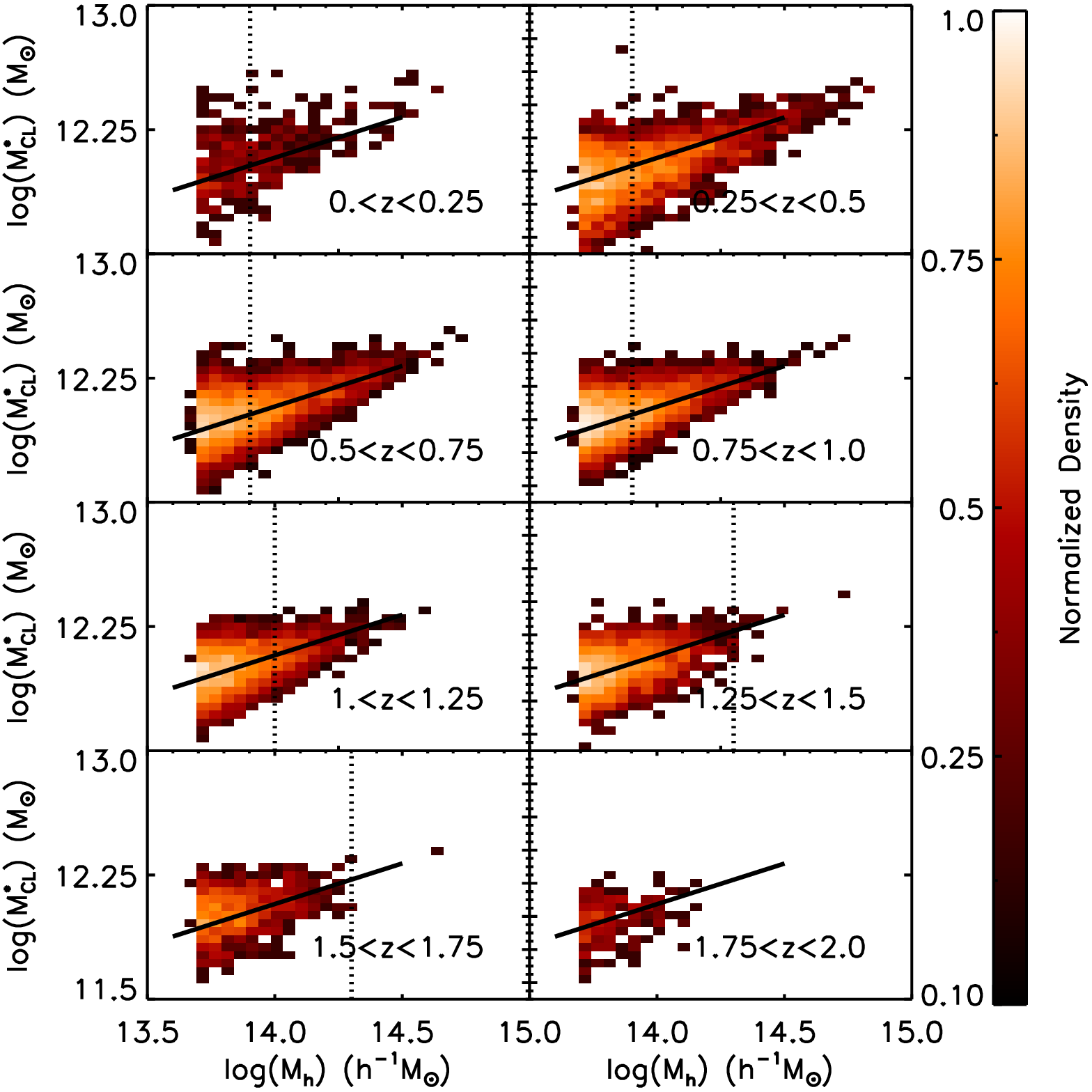} 
\caption{Density plots of the logarithm of the total stellar mass in the cluster as a function of the logarithm of the dark matter halo mass for the matched clusters in the Euclid-Pessimistic mock catalogue for different redshift bins. The solid line indicates the linear fit obtained down to $M_{\rm h}=7\times10^{13}  h^{-1} M_{\odot}$ for the entire redshift range. The vertical dotted line refers to the mass limit for which we can reliable detect galaxy clusters based on the analysis performed in section 5.2.}
\label{fig:smassmhaloEP}
\end{figure}

\begin{figure}
\centering
\includegraphics[clip,angle=0,width=1.0\hsize]{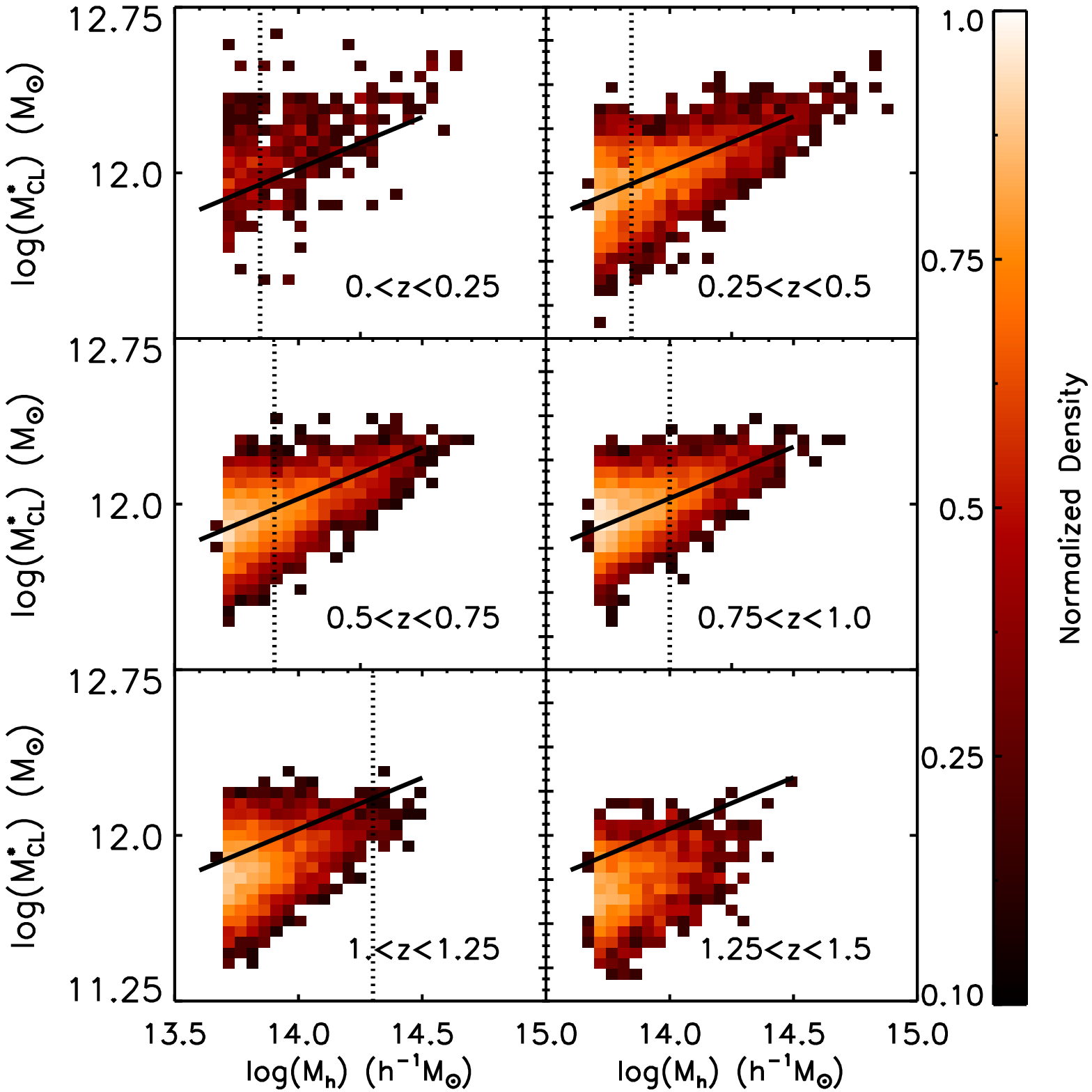} 
\caption{Density plots of the logarithm of the total stellar mass in the cluster as a function of the logarithm of the dark matter halo mass for the matched clusters in the LSST mock catalogue for different redshift bins. The solid line indicates the linear fit obtained down to $M_{\rm h}=7\times10^{13}  h^{-1} M_{\odot}$ for the entire redshift range. The vertical dotted line refers to the mass limit for which we can reliable detect galaxy clusters based on the analysis performed in section 5.2. We notice that at $z>1$, the fit does not represent the data due to incompleteness in our observable.}
\label{fig:smassmhaloL}
\end{figure}

We illustrate these results by showing density plots of $M^*_{\rm CL}$ versus halo mass for different redshift bins for the three different surveys (Euclid-Optimistic, Fig. \ref{fig:smassmhaloEO}; Euclid-Pessimistic Fig. \ref{fig:smassmhaloEP}; and LSST, Fig. \ref{fig:smassmhaloL}) together with the results of the fit (solid line). The vertical dotted line refers to the mass limit for which the completeness and purity rates are higher than 80\% according to the analysis performed in section 5.2. 

While the model fits the Euclid-Optimistic and Pessimistic surveys well (Fig. \ref{fig:smassmhaloEO} and Fig. \ref{fig:smassmhaloEP}), we see that it only describes the LSST data up to $z<1$; at higher redshift,  the fit becomes poorer. The most plausible explanation for this issue is that, at redshift $>1.0$, systematic errors in our observable, $M^*_{\rm CL}$,  are introduced due to the lack of the near-IR data for the LSST. While it exists the possibility that the redshift dependence in the model does not describe well the change in redshift, we should be able to see the same behaviour (a faster evolution in redshift) in the Euclid cases, and we do not. Future investigations will be devoted to consider a wider range of models to describe this relation.

The $M^*_{\rm CL} | M_{\rm h}$ relation  appears not to evolve significantly with redshift, in agreement with other works \citep{lin06,andreon14,saro15,ascaso16}. Nevertheless, \cite{vanuitert16} recently found evolution in the mass-richness relation, where their halo mass is estimated via weak lensing and the richness, $N_{200}$, is measured from the RCS2 survey \citep{gilbank11}. While the authors only consider two redshift bins with reliable mass measurements, different systematics regarding the selection of the sample and the aperture used to measure the richness might explain this difference. 

The scatter obtained for the $M^*_{\rm CL} - M_{\rm h}$ relation for the Euclid Optimistic ($\sim 0.124 \rm dex$),  Euclid Pessimistic ($\sim 0.135 \rm dex$) and LSST ($\sim 0.136 \rm dex$) surveys are comparable or smaller than those obtained observationally by other authors \citep{andreon10,andreon12,saro15} down to a similar mass threshold but extended to broader redshift ranges. This value is also similar to what was found in the J-PAS survey ($\sim 0.142 \rm dex$, \citealt{ascaso16}) which was limited to $z<0.7$, but reached masses $M_{\rm h}>5\times10^{13}  h^{-1} M_{\odot}$.

We have also performed the fit of the mass-observable relation, and estimated its corresponding scatter, $\sigma_{M_{\rm h}|M^*_{\rm CL} }$. In order to do this, we performed a similar approach to the one performed in \cite{ascaso15a,ascaso16}.  For each fixed value of $M^*_{\rm CL}$, we found the median and scatter values of the $M_{\rm h}$ by performing $10^4$ Monte Carlo samplings of all possible halo mass values.

The mean $\sigma_{M_{\rm h}|M^*_{\rm CL} }$ values obtained for the Euclid-Optimistic, Euclid-Pessimistic and LSST respectively are 0.196, 0.253 and 0.223 $\rm dex$, respectively if we consider matches with $M_{\rm h}>3\times10^{13}  h^{-1} M_{\odot}$. If we consider matches with $M_{\rm h}>1\times10^{13}  h^{-1} M_{\odot}$, these values increase by a factor of two. These  values are 2 times higher than the values found in \cite{ascaso15a,ascaso16} for the same mass limit for lower-z cluster samples. In addition, samples limited to  2-3 times higher masses found similar values \citep{rozo09,andreon12} or even lower  \citep{saro15}. This comparison is not straightforward due to the different cluster selection criteria and highlights the importance of performing consistent comparisons between different selection functions.

\subsection{Completeness and purity rates}
\label{ssec:secCP}

We have computed both completeness and purity rates as a function of redshift and mass for the matched list of detections.   We define completeness as the number of simulated cluster that have a counterpart on the cluster sample detected with the BCF without any cut restriction out of the total simulated sample. Similarly, we define purity as the number of clusters detected that have a counterpart in the original halo sample out of the total detected sample. In this work, we consider a cluster or group to have a halo mass $>3\times 10^{13} h^{-1} M_{\odot}$. Note that the fact that the catalogue is $\ge 16$ times smaller than the final area covered by the Euclid Wide / LSST survey is not important to these results. A smaller volume will miss the most massive clusters ($>1\times10^{15}M_{\odot}$), since they are the rarest (e.g. \citealt{warren06}).  However, these clusters are easy to detect by any methodology since their signal to noise is usually maximized, as discussed in other work (\emph{$A^2$}I, \citealt{ascaso16}). 

\begin{figure}
\centering
\includegraphics[clip,angle=0,width=1.0\hsize]{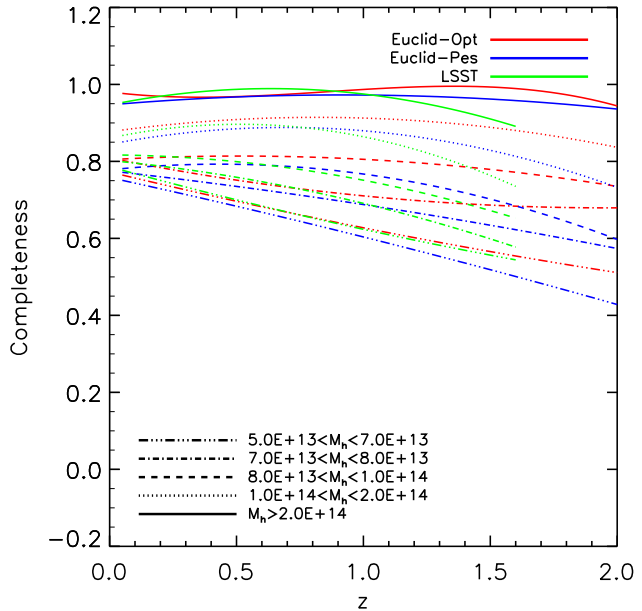} 
\caption{Completeness as a function of redshift for different dark matter halo mass ($M_{\rm h}(h^{-1}M_{\odot})$) bins for Euclid-Optimistic (red), Euclid-Pessimistic (blue) and LSST (green). The plotted lines have been smoothed by a second order polynomial  interpolation. These curves have been computed taking into account the observable-mass relation computed in section 5.1.  We note similar behaviour for the three surveys, with the Euclid-Optimistic completeness rates systematically higher than those for Euclid-Pessimistic and the LSST at any redshift and mass bin.}
\label{fig:completAll}
\end{figure}

\begin{figure}
\centering
\includegraphics[clip,angle=0,width=1.0\hsize]{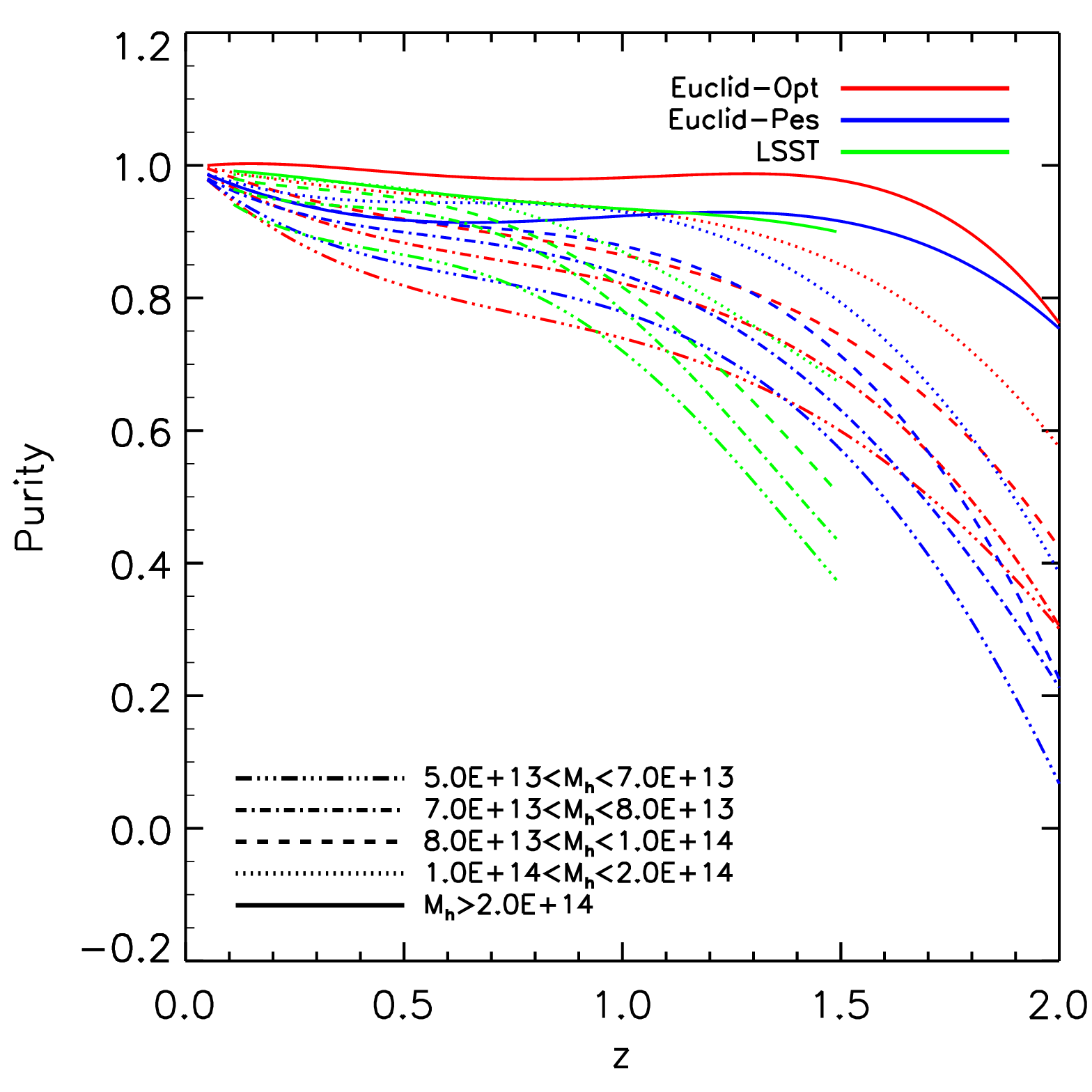} 
\caption{Purity as a function of redshift for different dark matter halo mass ($M_{\rm h}(h^{-1}M_{\odot})$)  bins for Euclid-Optimistic (red), Euclid-Pessimistic (blue) and LSST (green). The plotted lines have been smoothed by a fourth order polynomial  interpolation. These curves have been computed taking into account the observable-mass relation computed in section 5.1.  The purity rates remain $>80\%$ and are similar for the three surveys up to redshift $\sim 1$. At redshift $>1$ they decrease at different rates, with Euclid-Optimistic attaining the highest purity rates and LSST the lowest.}
\label{fig:purityAll}
\end{figure}

In Figs. \ref{fig:completAll} and \ref{fig:purityAll}, we show the completeness and purity rates for the three surveys for different halo mass $M_{\rm h}$ bins. The halo values for the  purity have been obtained after calibrating the $M^*_{\rm CL} - M_{\rm h}$ relation as described in section 5.1.  Tables \ref{tab:CompRates} and \ref{tab:PurityRates} summarize completeness and purity rates for four different redshift values.

\begin{table*}
      \caption{Completeness rates values corresponding to different halo mass bins and redshift values.}
      \[
         \begin{array}{c|ccccc|ccccc|ccccc}
		\hline
		{\rm z} &\multicolumn{5}{c}{ {\rm Euclid-Opt}} & \multicolumn{5}{c}{ {\rm Euclid-Pes}} & \multicolumn{5}{c}{ {\rm LSST}} \\\hline
		 &  {\rm M_1} & {\rm M_2}  & {\rm M_3}  & {\rm M_4} & {\rm M_5} &   {\rm M_1} & {\rm M_2}  & {\rm M_3}  & {\rm M_4} & {\rm M_5} &   {\rm M_1} & {\rm M_2}  & {\rm M_3}  & {\rm M_4} & {\rm M_5} \\\hline
0.5 &  0.70  & 0.75   & 0.81  & 0.91  & 0.97   & 0.68  & 0.74  & 0.79   & 0.88  & 0.97  & 0.70   & 0.76 &  0.80  & 0.90   & 0.98   \\
1.0 &  0.63  & 0.71   & 0.80  & 0.91  & 0.98   & 0.60  & 0.69  & 0.78   & 0.88  & 0.97  & 0.62   & 0.69 &  0.75  & 0.86   & 0.98   \\
1.5 &  0.57  & 0.69   & 0.78  & 0.89  & 0.99   & 0.52  & 0.63  & 0.70   & 0.83  & 0.96  & 0.57   & 0.60 &  0.67  & 0.76   & 0.91   \\
2.0 &  0.51  & 0.68   & 0.74  & 0.84  & 0.94   & 0.43  & 0.57  & 0.60   & 0.73  & 0.94  &  -   & - &  -  & -   & -   \\
\hline
	\end{array}
      \]
\begin{flushleft}
${\rm M_1}: 5\times10^{13}h^{-1} M_{\odot}<{\rm M_1} <7\times10^{13}h^{-1} M_{\odot}$, \quad ${\rm M_2}: 7\times10^{13}h^{-1} M_{\odot}<{\rm M_2} <8\times10^{13}h^{-1} M_{\odot}$, \quad ${\rm M_3}: 8\times10^{13}h^{-1} M_{\odot}<{\rm M_3 } <1\times10^{14}h^{-1} M_{\odot}$\\
${\rm M_4}: 1\times10^{14}h^{-1} M_{\odot}<{\rm M_4 } <2\times10^{14}h^{-1} M_{\odot}$, \quad ${\rm M_5}: {\rm M_5} >2\times10^{14}h^{-1} M_{\odot}$
\end{flushleft}
\label{tab:CompRates}
   \end{table*}

\begin{table*}
      \caption{Purity rates values corresponding to different halo mass bins and redshift values. The same redshift and halo mass bins as in Table \ref{tab:CompRates} have been used. }
      \[
         \begin{array}{c|ccccc|ccccc|ccccc}
		\hline
		{\rm z} &\multicolumn{5}{c}{ {\rm Euclid-Opt}} & \multicolumn{5}{c}{ {\rm Euclid-Pes}} & \multicolumn{5}{c}{ {\rm LSST}} \\\hline
		 &  {\rm M_1} & {\rm M_2}  & {\rm M_3}  & {\rm M_4} & {\rm M_5} &   {\rm M_1} & {\rm M_2}  & {\rm M_3}  & {\rm M_4} & {\rm M_5} &   {\rm M_1} & {\rm M_2}  & {\rm M_3}  & {\rm M_4} & {\rm M_5} \\\hline
0.5 &  0.82  & 0.88   & 0.92  & 0.96  & 0.99   & 0.85  & 0.90  & 0.92   & 0.94  & 0.92  & 0.86   & 0.93 &  0.95  & 0.97   & 0.96   \\
1.0 &  0.74  & 0.81   & 0.86  & 0.93  & 0.98   & 0.78  & 0.84  & 0.88   & 0.93  & 0.92  & 0.72   & 0.78 &  0.82  & 0.87   & 0.93   \\
1.5 &  0.60  & 0.68   & 0.74  & 0.85  & 0.98   & 0.57  & 0.63  & 0.71   & 0.79  & 0.92  & 0.37   & 0.43 &  0.50  & 0.67   & 0.90   \\
2.0 &  0.30  & 0.31   & 0.42  & 0.58  & 0.76   & 0.07  & 0.21  & 0.23   & 0.36  & 0.75  &  -   & - &  -  & -   & -   \\
\hline
	\end{array}
      \]
\label{tab:PurityRates}
   \end{table*}

 The completeness shows a similar behaviour for the three surveys, with high ($\sim 90\%$) completeness rates at the high-mass end ($>2\times 10^{14} h^{-1} M_{\odot}$). At smaller masses, the completeness decreases as a function of redshift, with the fastest decline occurring in the lowest mass bins. We also note that the completeness rates for the Euclid-Optimistic case are higher than those for the Euclid-Pessimistic and the LSST cases for the same mass bins at redshift $> 0.5$. The completeness rates for the LSST and DES are  very similar up to redshift $\sim 1.2$, while at higher redshifts the rates for the LSST become smaller than those for the Euclid-Pessimistic.

 The differences in purity are larger than those for completeness. As expected, the three surveys show decreasing purity rates as a function of redshift for a fixed halo mass and also towards lower masses. Up to redshift $\sim 1$, all the purity rates remain $>80\%$ for any halo mass bin. Moreover, the differences are very small between the three surveys up to $z\sim1$, at least for clusters more massive than $7\times 10^{13} h^{-1} M_{\odot}$. At higher redshifts, the LSST purity rates decrease the fastest with redshift, while the purity rates for the Euclid-Optimistic survey decrease the slowest.

As we see, the different properties of the surveys considered in this work in terms of depth and photometric redshift resolution affect the expected cluster sample, and therefore their selection functions.  We explore the selection functions derived from these rates and the observable-mass relation in the next section.

\subsection{Selection Function}

In this section, we compute the selection function for the three surveys considered in this work. We first use the completeness and purity rates obtained section 5.2, to obtain the minimum halo mass for which both quantities are higher than a particular threshold at a particular redshift. The purity curves, originally as a function of the observable mass, have been calibrated using the  observable-mass relation and its scatter computed in section 5.1 down to this minimum halo mass. We then define the selection function as the minimum mass that we are able to detect with  higher completeness and purity rates than a particular limit as a function of redshift.

\begin{figure}
\centering
\includegraphics[clip,angle=0,width=1.0\hsize]{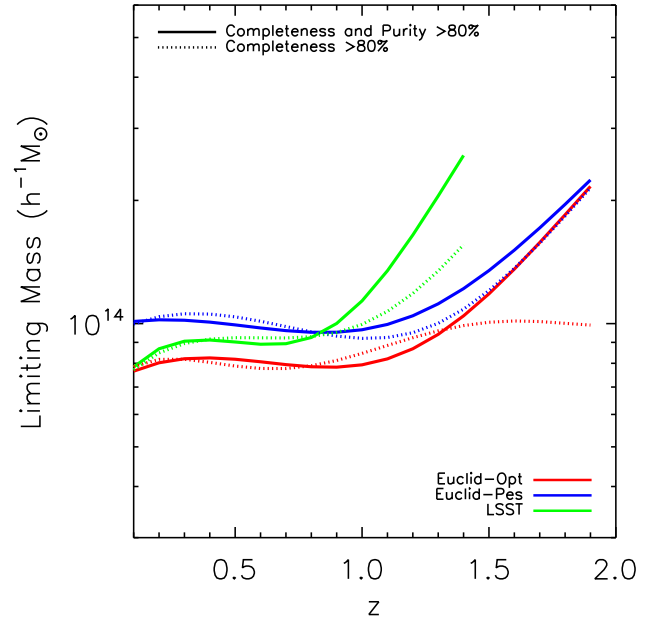} 
\caption{ Selection functions (minimum mass threshold as a function of redshift) obtained from different threshold limits in completeness and purity rates for Euclid-Optimistic, Euclid-Pessimistic and LSST (red, blue and green lines respectively). The solid lines refer to the selection functions obtained assuming both completeness and purity rates higher than 80\% and the dotted lines refer to the selection functions obtained assuming only completeness rates higher than 80\%.}
\label{fig:selectionComp}
\end{figure}

The selection function depends on the thresholds applied to both completeness and purity. In Fig. \ref{fig:selectionComp}, we display the selection functions obtained with a threshold of 80\% on both completeness and purity (solid lines) and  with a threshold of 80\% on completeness alone (dashed lines). At low redshift, the selection functions obtained by imposing thresholds on both completeness and purity are similar to those when imposing a threshold on completeness alone ($z<1.4$, $z<1$ and $z<0.8$ for Euclid-Optimistic, Euclid-Pessimistic and LSST respectively). At higher redshift, the two sets of curves deviate, with the completeness only curves remaining flatter.  Note that these curves have been smoothed by interpolation, which also introduces an uncertainty between  7 and 10\%.  The impact of this error is moderate and explains visual effects such as the dotted curves rising above the solid ones in few cases.

We have compared our empirical Euclid cluster selection functions with the ones used by  \cite{sartoris16}  based on an analytical approximation of cluster detection at 3 and 5$\sigma$ expressed in units of Dhalo masses \citep{jiang14}. Their 3$\sigma$ selection function agrees within the errors with our overall Euclid-Optimistic scenario when we impose only a completeness threshold. When we add the additional constraint on purity, our selection function steepens at $z>1.5$. This is very  good agreement given the use of the different methodologies.

\begin{figure}
\centering
\includegraphics[clip,angle=0,width=1.0\hsize]{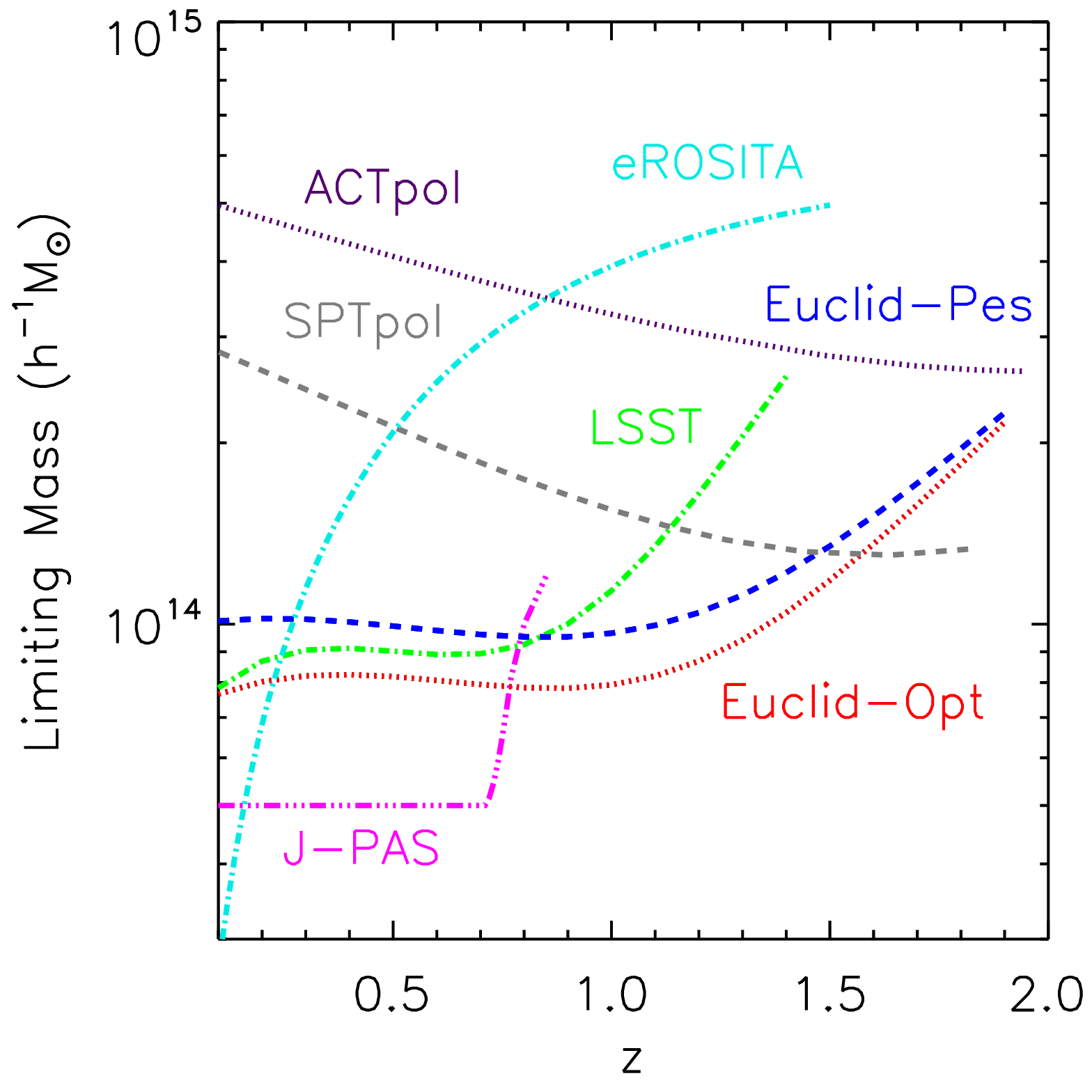} 
\caption{Selection function (minimum mass threshold as a function of redshift) for different next-generation surveys. The three surveys considered in this work are represented with a red dotted line (Euclid-Optimistic), a blue dashed line (Euclid-Pessimistic) and a green dotted-dashed line (LSST)  and they are obtained by imposing both completeness and purity rates $>80\%$. For comparison we have displayed other next-generation surveys: the J-PAS optical survey (pink three dotted-dashed line); the eROSITA X-ray survey (cyan dotted-dashed line) and the ACTpol and SPTpol SZ survey (purple dotted and dashed gray lines respectively). The J-PAS selection function has been obtained from Ascaso et al. 2016,  the X-ray selection function from Pillepich et al. 2012, and the SZ selection functions from Weinberg et al. 2013.}
\label{fig:selectionF}
\end{figure}

In Fig. \ref{fig:selectionF}, we show the selection function of the three surveys. For comparison, the J-PAS selection function, obtained in the same way as this work \citep{ascaso16}, is shown along with the selection functions from X-ray (eROSITA,  \citealt{pillepich12}) and SZ surveys (ACT or SPT, \footnote{Extracted with Dexter, http://dexter.sourceforge.net/} \citealt{weinberg13}).  The eROSITA curve was originally expressed as a function of $M_{500}$ so we have used the prescription given by \cite{hu03} to translate those masses into virial masses to perform a more consistent comparison.

If we first focus on the four different optical surveys displayed in Fig. \ref{fig:selectionF}, we can see that all of them show similar behaviour. They are almost flat at lower redshift, and they progressively increase at higher redshift. We also notice that the selection functions of Euclid-Optimistic and Euclid-Pessimistic surveys have very similar shape, with $\sim 10\%$ shift in mass. This shift is caused by the better accuracy on the photometric redshift in the former survey leading to higher purity rates. Furthermore,  the LSST selection function reaches $\sim 6\%$ lower masses than Euclid-Pessimistic up to $z\sim0.7$, which indicates the benefits of using deeper magnitudes and a larger number of bands in the optical to obtain lower detection mass thresholds. In this respect, the photometric redshift scatter is directly proportional to the minimum mass threshold that we can resolve, as  has been noted by \cite{ascaso15a} and \cite{ascaso16}. In this plot, it becomes clear when comparing the LSST and J-PAS selection functions at $z<0.7$. J-PAS, with more than 50 narrow-bands, has a higher photometric redshift accuracy that allows it to sample  at least $\sim 1.5$ times lower in mass with respect to LSST. 

Complementarily, the depth of the survey plays a crucial role in the optical selection function. Very deep surveys such as LSST sample lower masses than DES. Also, very deep data benefits from better photometric redshifts at high redshift, as can be seen in the Euclid-Optimistic survey. The Euclid survey with deep IR bands also allows exploration of a broader range of redshift, $z<2$, sampling at least the more massive end of the mass function up to this redshift.

Interestingly, the selection functions obtained by different methodologies have very different shapes. The X-ray selection function displays a progressive increase of the limiting mass with redshift, showing smaller limiting masses than the optical surveys at very low redshifts ($z<0.2-0.3$). In contrast, the SZ selection functions show a progressive decrease with redshift  achieving mass thresholds comparable to those found in optical surveys at redshifts higher than 1.1 for the LSST and 1.5 for the Euclid surveys.

The complementarity of the different methodologies is clear. Different surveys will map different clusters having a variety properties. The combination of all these methods will provide a complete view of the properties of  the overall cluster population.

\subsection{Accuracy on the cluster redshift estimation}

In \emph{$A^2$}I, we investigated in detail the photometric redshift performance of individual galaxies for the three surveys. The Euclid-Optimistic survey showed a factor of $\ge 2$ lower scatter and outlier rate than Euclid-Pessimistic and LSST for a wider range of redshift and mass. This is expected since the number of bands covering the optical and IR spectrum is maximized for this survey. Moreover,  the photometric redshift performance in the LSST survey was slightly better than for the Euclid-Pessimistic survey up to $z<1$.  However, at higher redshift the LSST dispersion and outlier rates increased and departed significantly from the Euclid-Pessimistic case. 

In this section, we consider the accuracy of cluster redshift recovery for the different surveys. To do this, we have used the NMAD estimator of the dispersion  (\citealt{brammer08}, see also \citealt{ascaso15b,ascaso16}), as it is known to be a robust representation of the distribution. We define the NMAD estimator as:

\begin{equation}
\sigma_{\rm NMAD}=1.48\times {\rm median}\bigg(\frac{|\Delta z - {\rm median}(\Delta z)|}{1+z_{\rm s}}\bigg)
\end{equation}
where $z_{\rm s}$ is the spectroscopic redshift, $\Delta z=z_{\rm CL}-z_{\rm s}$ and $z_{\rm CL}$ is the estimated cluster redshift.

\begin{figure}
\centering
\includegraphics[clip,angle=0,width=1.0\hsize]{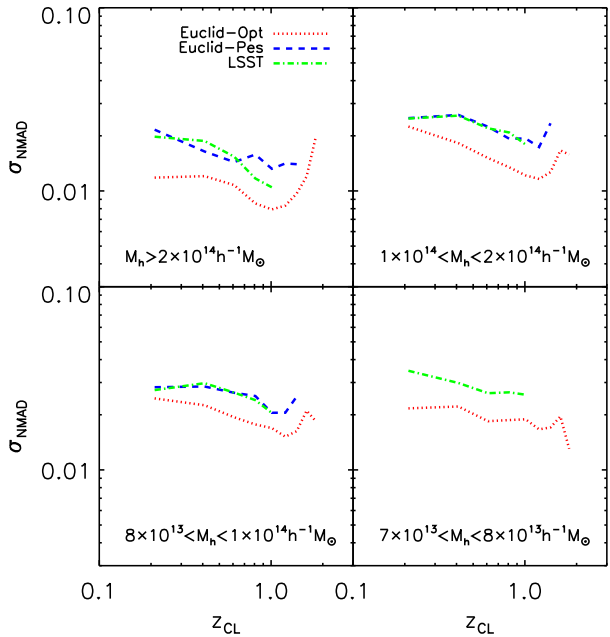} 
\caption{Dispersion between the estimated cluster redshift and the spectroscopic redshift of the cluster as a function of redshift for different halo mass slices ($>2\times10^{14}  h^{-1} M_{\odot}$, top left panel, ($1\times10^{14}  h^{-1} M_{\odot}<M_{\rm h}<2\times10^{14}  h^{-1} M_{\odot}$, top right panel, ($8\times10^{13}  h^{-1} M_{\odot}<M_{\rm h}<1\times10^{14}  h^{-1} M_{\odot}$, bottom left panel, and ($7\times10^{13}  h^{-1} M_{\odot}<M_{\rm h}<8\times10^{13} h^{-1}  M_{\odot}$, bottom right panel) for the Euclid-Optimistic (red dotted line), Euclid-Pessimistic  (blue dashed line) and the LSST  (green dashed-dotted line) surveys.}
\label{fig:clustersz}
\end{figure}

In Fig.~\ref{fig:clustersz}, we display the average NMAD dispersion between the photometric redshift estimate and the spectroscopic redshift  as a function of redshift for different halo mass thresholds in the three surveys.

As expected, the quality of the individual photometric redshift of the galaxies has a direct impact on the photometric redshift recovery of the cluster. The Euclid-Optimistic survey has the lowest scatter for any mass threshold and at any redshift, whereas the Euclid-Pessimistic survey has almost a factor of two higher scatter, similar to what was found for individual galaxies. We do not observe any significant dependence of the redshift dispersion on redshift. For each survey, they remain almost constant for any mass threshold.

We also show that the most massive ($M_{\rm h}>2\times10^{14} h^{-1} M_{\odot}$) clusters recover the expected cluster redshift with lower dispersion ($\sigma_{\rm NMAD} \sim 0.01$ for the Euclid-Optimistic, $\sigma_{\rm NMAD} \sim 0.015$ for the Euclid-Pessimistic and LSST) than for lower masses thresholds. This is also expected since the clusters have more member galaxies, and therefore the estimated redshift is more robustly determined.

\begin{figure}
\centering
\includegraphics[clip,angle=0,width=1.0\hsize]{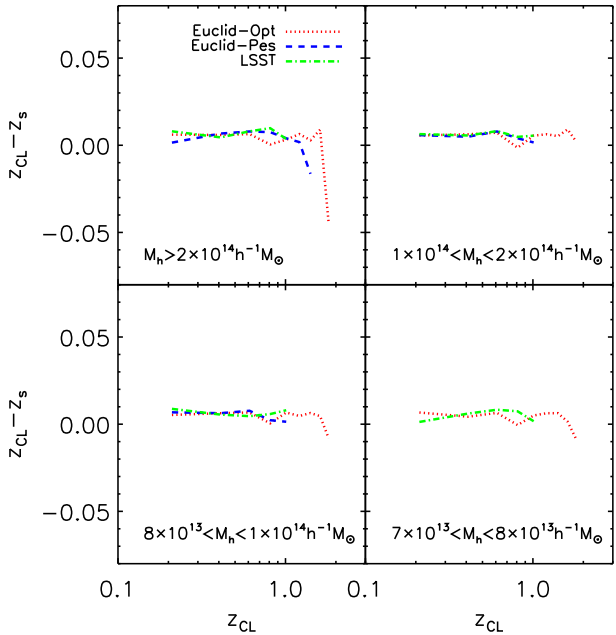} 
\caption{Difference between the estimated cluster redshift and the spectroscopic redshift of the cluster as a function of redshift for different halo mass slices ($>2\times10^{14}  h^{-1} M_{\odot}$, top left panel, ($1\times10^{14}  h^{-1} M_{\odot}<M_{\rm h}<2\times10^{14}  h^{-1} M_{\odot}$, top right panel, ($8\times10^{13}  h^{-1} M_{\odot}<M_{\rm h}<1\times10^{14}  h^{-1} M_{\odot}$, bottom left panel, and ($7\times10^{13}  h^{-1} M_{\odot}<M_{\rm h}<8\times10^{13}  h^{-1} M_{\odot}$, bottom right panel) for the Euclid-Optimistic (red dotted line), Euclid-Pessimistic  (blue dashed line) and the LSST  (green dashed-dotted line) surveys.}
\label{fig:clustersbiasz}
\end{figure}

Similarly, in Fig.  \ref{fig:clustersbiasz}, we display the redshift bias, $\Delta z$, i.e. the difference between the cluster photometric redshift estimation and the spectroscopic redshift of the considered cluster, as a function of redshift for different halo mass thresholds for the three surveys.  The three surveys show negligible bias, confirming  the reliability of the detections. In addition, they do not display any dependence of the bias as a function of redshift.

We then conclude that both the scatter and the bias in the recovery of the cluster redshift is within the photometric redshift dispersion of the survey galaxies, indicating that the cluster redshift can be determined with an accuracy better than the individual galaxies.

\section{Conclusions}

This work, the second of the `Apples to Apples' \emph{$A^2$} series, aims at consistently comparing the  selection function and mass-observable relation obtained for the optical  galaxy cluster catalogues expected from  two different next-generation surveys:  LSST and Euclid. The expected cluster and group catalogue have been obtained by detecting them with the Bayesian Cluster Finder (BCF, \citealt{ascaso12,ascaso14a}) in mock catalogues mimicking the two surveys. In the previous work of the series, \emph{$A^2$}I, we characterized these mock catalogues, ensuring the similarity of their properties with the real universe.

The mass observable adopted in this work is the total stellar mass, $M^*_{\rm CL}$. We modeled the observable-mass $M^*_{\rm CL}-M_{\rm h}$ relation as a power law, including a redshift dependence term and fit it to the different datasets. The results of the fit are consistent with no evolution in the relation with redshift. However, for the LSST case, we have  restricted our fit to $z<1$, since at higher z the measurements are not reliable due to the errors in the mass measurement. The non-evolution of the relation also agrees with different empirical and theoretical results in the literature \citep{lin06,andreon14,saro15,ascaso16}. However, recent results by \cite{vanuitert16} find evolution within the redshift range $0.2\le z \le 0.55$, at least. This discrepancy might arise from the differences in the selection of the sample and the computation of the richness.

Furthermore, we have obtained values for the scatter of $\sigma_{M^*_{\rm CL}|M_{\rm h}} \sim$ 0.124, 0.135 and 0.136 $\rm dex$, for Euclid-Optimistic, Euclid-Pessimistic and LSST, respectively. These values are comparable to those from other studies analyzing limited samples of clusters \citep{andreon10,andreon12,saro15} restricted to a narrower redshift range; and they are slightly smaller than those of the J-PAS survey \citep{ascaso16}, for  $z<0.7$ clusters and extending over a larger range in mass ($M_{\rm h}>5\times10^{13} h^{-1} M_{\odot}$).  This result highlights the potential of the Euclid surveys to measure the masses of galaxy clusters at high redshift with accuracies comparable to those at low redshift.

 Similarly, we have computed  the scatter in the mass-observable relation $\sigma_{M_{\rm h}|M^*_{\rm CL}}$, finding 0.196, 0.253 and 0.223 $\rm dex$, for Euclid-Optimistic, Euclid-Pessimistic and LSST, respectively, considering clusters matched down to $M_{\rm h}=3\times10^{13} h^{-1} M_{\odot}$. These values are similar to those from other studies based on simulations (e.g. \citealt{ascaso16}), but are larger than results from observational samples restricted to higher  ($>2\times10^{14}M_{\odot}$) masses (e.g \citealt{rozo09,andreon12,saro15}), which however also makes direct comparison difficult.

 We have furthermore computed completeness and purity rates for the three surveys. We found completeness rates at $>80\%$ for clusters with  $M_{\rm h}>1\times10^{14} h^{-1} M_{\odot}$ up to $z \sim 2$, $1.7$ and $1.3$, for Euclid-Optimistic, Euclid-Pessimistic and LSST, respectively, and $M_{\rm h}>8\times10^{13} h^{-1} M_{\odot}$ up to $z \sim 0.5$, for Euclid-Optimistic, and LSST. The purity rates become $> 80\%$ for $M_{\rm h}>1\times10^{14} h^{-1} M_{\odot}$ up to $z \sim 1.6$, $1.5$ and $1.2$ for Euclid-Optimistic, Euclid-Pessimistic and LSST, respectively, and for $M_{\rm h}>8\times10^{13} h^{-1} M_{\odot}$ up to $z \sim 1.3$, $1.3$ and $1.0$, for the same surveys.

Using the fits to the $M^*_{\rm CL}-M_{\rm h}$ relation and the expected completeness and purity rates, we computed selection functions for the different surveys.  We compared our selection functions obtained when imposing an 80\% threshold on both completeness and purity to those obtained when imposing a threshold of 80\% on completeness alone, finding that the latter results in flatter selection functions at higher redshift ($z>1.4$ for Euclid-Optimistic, $z>1$ for Euclid-Pessimistic and $z > 0.8$ for LSST).  This latter selection function agrees with the one used by \cite{sartoris16}, an encouraging agreement given the different methodologies employed.

We compared our selection functions at a threshold of 80\% in both completeness and purity to those from other  works using optical and non-optical data. Note that these selection functions are the first ones obtained for these surveys by using an empirical detection of clusters in simulations. We find that the shapes of the selection functions in the optical are very similar among themselves. The Euclid-Optimistic and Euclid-Pessimistic surveys are able to detect galaxy clusters up $z<2$ at least down to $M>2\times 10^{14} h^{-1} M_{\odot}$. The Euclid-Optimistic shows a 13\% difference in normalization  in mass with respect to the Euclid-Pessimistic survey. The selection function of LSST samples $\sim 6\%$ lower masses than Euclid-Pessimistic up to $z\sim0.7$. At higher redshift, LSST increases in limiting mass faster than the Euclid surveys, due to the absence of IR data.  Similarly, the selection function of J-PAS \citep{ascaso16} samples 38.5\% lower masses than Euclid-Optimistic up to $z\sim0.7$.

We compared these optical selection functions to selection functions applicable to other survey methods. While the X-ray eROSITA selection function also increases with redshift, the SZ SPTpol and ACTpol selection functions decrease.  Closer inspection highlights their complementarity.  The main cluster and group budget down to masses $<10^{14} h^{-1} M_{\odot}$ will be mapped by eROSITA at redshift $<0.3$. LSST and Euclid will detect clusters and groups down to $8\times10^{13}-2\times10^{14} h^{-1} M_{\odot}$ within the redshift range $0.3\le z \le 1.1$ and $0.3\le z \le 1.5$, respectively. Finally, higher redshift ranges $z>1.5$ will be accessible with SPTpol down to masses of $2\times10^{14} h^{-1} M_{\odot}$ or even lower at higher redshifts. The combination of these different survey methods will provide a more complete and robust view of cluster properties, at least up to $z<1.5$, and therefore more robust cosmological constraints from the counts.

In a forthcoming paper of the series, \emph{$A^2$}III, we will explore the translation of these selection functions into predictions for cosmological constraints using cluster counts.

\section*{Acknowledgments}

We acknowledge the anonymous referee for providing useful comments that helped improving this manuscript. BA thanks Thomas Reiprich and Annalisa Pillepich, who kindly provided the updated e-ROSITA selection function curves. This project has received funding from the European Union's Horizon 2020 research and innovation programme under the Marie Sklodowska-Curie grant agreement No 656354 and support from a postdoctoral fellowship at the Observatory of Paris. S.M. acknowledges financial support from the Institut Universitaire de France (IUF), of which she is senior member.  We thank Peter Schneider and Andrea Biviano for useful comments on the manuscript.

\end{document}